# Evaluation on characterization of acoustic emission of brittle rocks from the experiment to numerical simulation

Fengchang Bu[1,2,3], Lei Xue[1,2,3]*, Mengyang Zhai[1,2,3], Xiaolin Huang[1,2,3]*, Jinyu Dong[4], Ning Liang[1,2,3], Chao Xu[1,2,3]

1. Key Laboratory of Shale Gas and Geoengineering, Institute of Geology and Geophysics, Chinese Academy of Sciences, Beijing 100029, China
2. Innovation Academy for Earth Science, Chinese Academy of Sciences, Beijing 100029, China
3. College of Earth and Planetary Sciences, University of Chinese Academy of Sciences, Beijing 100049, China
4. Research Institute of Geotechnical Engineering and Hydraulic Structure, North China University of Water Resources and Electric Power, Zhengzhou 450046, China

*Corresponding authors:

xuelei@mail.iggcas.ac.cn (Lei Xue)

huangxiaolin@mail.iggcas.ac.cn (Xiaolin Huang)

# Abstract

Acoustic emission (AE) characterization is an effective technique to indirectly capture the progressive failure process of the brittle rock. In previous studies, both the experiment and numerical simulation were adopted to investigate AE characteristics of the brittle rock. However, as the most popular numerical model, the moment tensor model (MTM) did not reproduce the monitoring and analyzing manner of AE signals from the physical experiment. Consequently, its result could not be constrained by the experimental result. It is thus necessary to evaluate the consistency and compatibility between the experiment and MTM. To fulfill this, we developed a particle-velocity-based model (PVBM) which enabled directly monitor and analyze the particle velocity in the numerical model and had good robustness. The PVBM imitated the actual experiment and could fill in gaps between the experiment and MTM. AE experiments of Marine shale under uniaxial compression were carried out, of which results were simulated by MTM. In general, the variation trend of the experimental result could be presented by MTM. Nevertheless, magnitudes of AE parameters by MTM presented notable differences with more than several orders compared with those by the experiment. We sequentially used PVBM as a proxy to analyze these discrepancies quantitatively and make a systematical evaluation on AE characterization of brittle rocks from the experiment to numerical simulation, considering the influence of wave reflection, energy geometrical diffusion, viscous attenuation, particle size as well as progressive deterioration of rock material. It was suggested that only the combination of MTM and PVBM could reasonably and accurately acquire AE characteristics of the actual AE experiment of brittle rocks by making full use of their respective advantages.

**Keywords:** Brittle rock; Acoustic emission experiment; Moment tensor model; Particle-velocity-based model; Robustness; Consistency and compatibility evaluation

# 1 Introduction

The failure of the brittle rock is a progressive process, which experiences crack initiation, crack propagation, and crack coalescence in turn[1,2]. By monitoring and modeling this process, engineers can conduct protection and prevention of rock mass engineering disasters such as the slope stability and rock burst[3-6]. However, it is always difficult to directly observe the progressive process of the rock failure for the practical application. Fortunately, the acoustic emission (AE) phenomenon accompanies with the fracturing event, of which characteristics can indirectly reflect the progressive failure process of the brittle rock[7-9]. Hence, it is of great significance to capture AE characteristics of brittle rocks for the monitoring and warning of the rock mass engineering disaster[10-13].

Laboratory observations were first conducted to investigate AE characteristics of brittle rocks when loaded. Elastic waves from the fracturing event were recorded by the piezoelectric sensor attached to the surface of the rock sample as shown in Fig. 1a. Then, these signals were quantitively analyzed to determine the AE count, energy release and the *b*-value, which presented the progressive failure process of the brittle rocks. However, there remain two main limitations for the experimental method[14]. On the one hand, only when the AE signal propagated to the sample surface, could it be monitored by piezoelectric sensors. During this process, the radiated energy was attenuated due to both geometrical diffusion and viscous dissipation. The monitored signals thus could not completely represent in-situ cases. On the other hand, the monitored waves were superimposed waves resulted from multiple wave reflections at interfaces, which posed challenges to recognize a series of independent AE events. Meanwhile, the progressive failure of the stressed brittle rocks could be only represented as AE signals monitored by a limited number of sensors. Also, the fracturing process inside the rock sample could not be visible in a real time to calibrate the AE signals.

Out of the experiment, the numerical model was also applied, which had the capability of simultaneously characterizing the rock failure and AE behaviors. Via the finite element method (FEM), the fracturing behavior is usually achieved by degrading material properties according to continuum laws and a damaged element is regarded as an acoustic event, which was reflected in a typical code called RFPA proposed by Tang[15]. Based on a similar principle, researchers have developed some other AE simulation models such as local linear tensorial damage model[16], local degradation model[17] and elasto-plastic cellular automata model[18]. Using the discrete element method (DEM), a rock material is expressed in the form of discrete elements connected by contact bonds. It makes quasi-static deformation simulation possible by solving a motion equation. Typically, Hazzard and Young[19] proposed a dynamical AE recording technique by directly quantifying the kinetic energy of particles into the energy radiated by seismic sources when bonds are broken in a 2D particle-based model. Then this technique was improved by introducing moment tensor calculation by tracking the change of contact force at the time of bonds broken, which is referred to as the moment tensor model (MTM)[20]. Besides FEM and DEM, there are some other AE simulation models such as static lattice model[21], continuum fracture mechanics model[22], quasi-dynamic monitoring kinetic energy model[23,24], deviatoric strain rate model[25] and Voronoi element based explicit numerical manifold model[26]. Among the models aforementioned, MTM based on DEM has been widely used in simulating AE benefiting from the ability of DEM to explicitly represent fractures and bond failure of rocks and excellent applicability of MTM to quantitative characterization of AE[27-30].

On the above basis, many researchers attempted to explore the AE characteristic of the brittle rock from the experiment to MTM directly. Ma et al. acquired different numbers of AE hits compared with experimental results by reproducing Brazilian tests in MTM[31]. Chorney et al. compared MTM with experiments in AE energy by simulating triaxial compression tests of sandstone[32]. Zhang and Zhang investigated the difference in the relative magnitude of *b*-value drop-offs between experiments and MTM by modeling uniaxial compression tests of limestone[33]. However, it is not convincing to directly compare simulated AE characteristics with those in experiments due to some key contradictories between MTM and the actual AE test[34] as shown in Fig. 1. In terms of principle, MTM treats bond breakages occurring close in time and space as a single AE hit (Fig. 1e) while actual AE tests treat superposed elastic waves that exceed a threshold and cause a system channel to accumulate data (Fig. 1b) as a single AE hit [20,35]. In terms of processing method, MTM realizes quantitative AE characterization by calculating moment tensor based on the change in contact forces upon particle breakage (Fig. 1e) while actual AE tests quantify AE by analyzing superimposed waveforms (Fig. 1b) acquired from AE sensors arranged on a surface of specimens[36,37]. In terms of energy analysis, in-situ energy can be explicitly calculated by MTM (Fig. 1f). While energy by the experiment is just a part of in-situ energy because of geometrical diffusion and viscous dissipation from the epicenter to the AE sensor location (Fig. 1c). It is worth noting that the potential breakable bonds in the numerical model are often far less than those of the actual rock considering the computational capability, resulting in smaller numbers of AE events by MTM than the experiment, especially for the 2D numerical model.

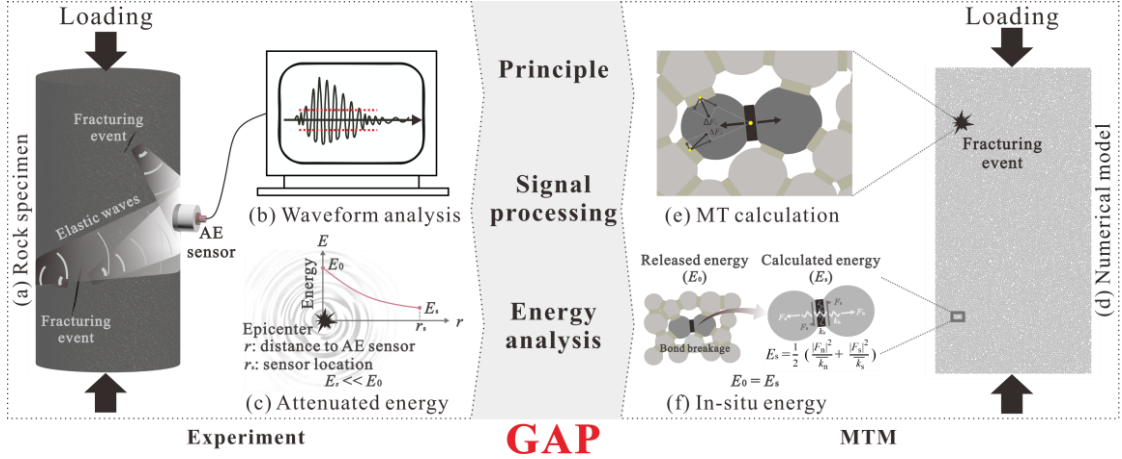

**Figure 1.** Schematic view on AE characterization by the experiment and MTM

From literature review above, it can be seen that the result by MTM cannot be constrained by the experiment result as it did not reproduce the monitoring and analyzing manner of AE signals of the physical experiment. It is thus important to evaluate the consistency and compatibility between the experiment and MTM, which is the motivation of this study. This paper was structured as follows. Section 1 reviews the previous experimental and numerical studies on the AE characteristics of the brittle rocks and analyzes the limitations in these studies. Section 2 overviews the MTM and simulates an AE experiment. Section 3 introduces the proposed PVBM and shows reproduced results using PVBM in detail. These results are discussed in Section 4. Conclusion is given in Section 5.

## 2 AE characterization by the experiment and MTM

### 2.1 Experiment

The Marine shale cylindrical specimens cored from the Longmaxi Formation in the Pengshui shale gas area in China were used to conduct uniaxial compression tests with AE in a laboratory. As shown in Fig. 2a, the samples all were applied to unconfined compressive load to failure with a constant rate of $10^{-5}s^{-1}$ by the RTR-2000 tri-axial dynamic testing system of GCTS (USA). Figure 2b shows the AE monitoring system with six AE sensors mounted on the surface of the specimen by rubber band and tape. The sampling frequency was set as 1MHz and the amplitude threshold was set as 35dB. Actual AE parameters including hit, energy and *b*-value were acquired by this system.

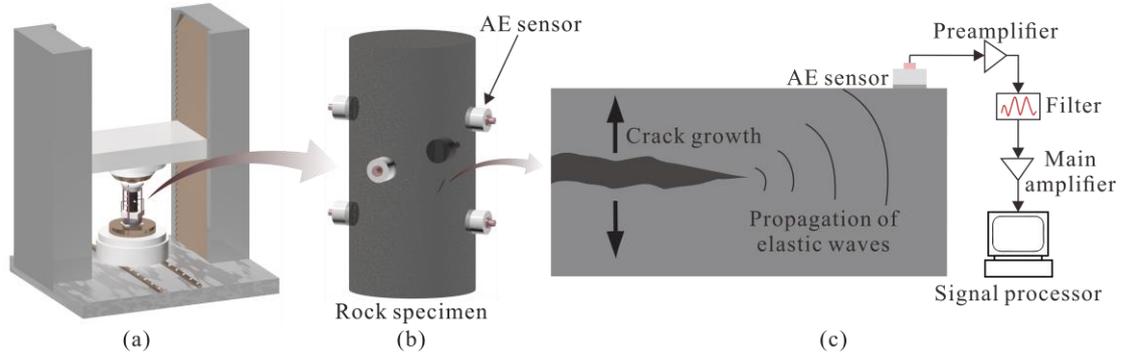

**Figure 2.** Schematic view on uniaxial compression tests with AE recording. (**a**) Loading system; (**b**) AE monitoring system; (**c**) AE signal processing

Figure 3 shows the variation of axial stress, cumulative number of cracks, and actual AE parameters (cumulative energy, *b*-value, and AE hits) versus axial strain. At the beginning of loading (strain from 0 to 0.06%), the stress-strain relation exhibits an approximately linear manner. Note that curves of cumulative energy and cracks increase slightly with the strain increasing which corresponds to some AE hits. The *b*-value is initially about 1.54 ± 0.26. The above phenomena may be resulted from the compaction of natural cracks in the shale specimen or the rupture of small bulges at the ends of the specimen[38]. As the strain ranges from 0.06% to 0.26%, the stress curve increases almost linearly while curves of cumulative energy and cracks hardly increase, involved in few AE hits. At this stage, the *b*-value curve basically keeps invariant, which indicates that the magnitude distribution of AE event is stable in general. These phenomena show that the specimen nearly produced a linear elastic response at this stage. When the strain falls in the range (0.26% - 0.69%), the stress curve increases nonlinearly and curves of cumulative energy and cracks increase greatly, with respect to approximately linear growth of AE hits. The *b*-value curve increases gradually, then it fluctuates and occurs a sudden drop when the strain reaches 0.67%. Interestingly, AE hits increase approximately in a linear manner with strain ranging from 0.33 % to 0.52%, out of which it is almost stable at 600 until the strain of 0.62%. Then AE hits drop sharply and the variation rate of cumulative energy and cracks decreases, suggesting that the rock might experience a transition from crack stable growth to unstable growth. Finally, when the strain is over 0.69%, the stress curve lies in postpeak stages, and curves of cumulative energy increase fiercely, corresponding to the strong increase of AE hits. The *b*-value curve drops significantly, which indicates the specimen reaches peak strength (unconfined compression strength, UCS), the cracks coalesce with each other on a large scale. Sequentially, the whole specimen was fractured into some pieces.

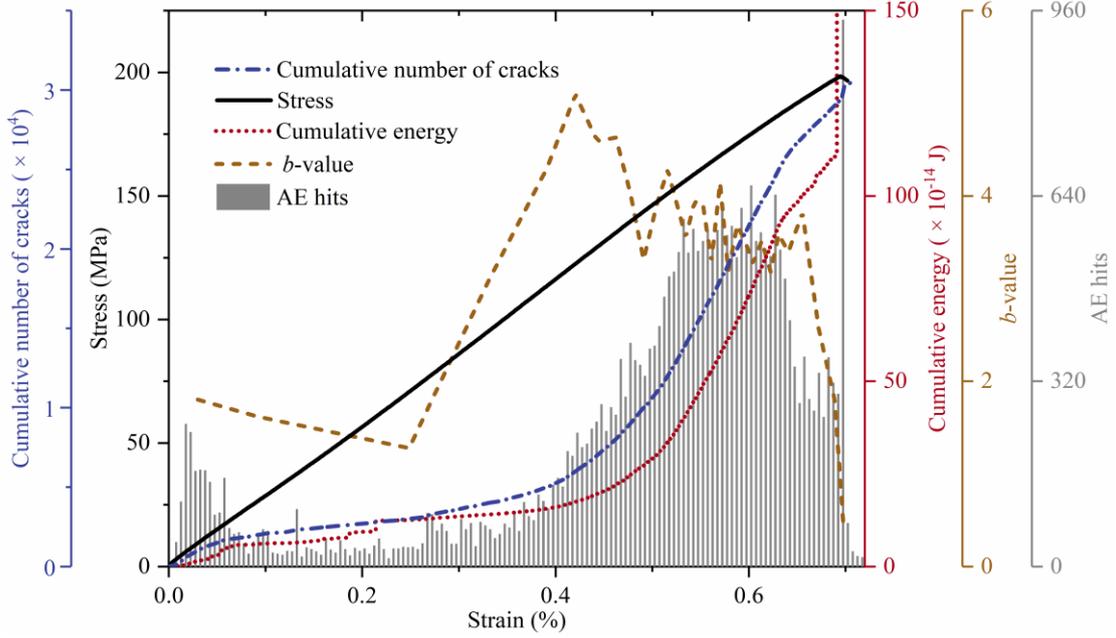

**Figure 3.** Axial stress, cumulative number of cracks and AE parameters (cumulative energy, *b*-value and AE hits) versus the axial strain by an actual AE test

## 2.2 MTM

### 2.2.1 Theory of MTM

MTM algorithm is implemented based on Particle Flow Code (PFC) of the DEM developed by the ITASCA Consulting Group (USA), of which a rock material is modeled as an assembly of bonded particles[19,39]. When external loads exceed the bond strength between two particles, it will be broken and release stored energy, causing movement of particles on both sides and deformation of adjacent contact bonds[19,40,41]. These changes can be quantified by calculating components of the moment tensor by summing the force change at the contact multiplies the distance of the contact from the bond breakage location[20]:

$$M_{ij} = \sum \Delta F_i R_j \qquad (1)$$

where $M_{ij}$ is the value of moment tensor, $\Delta F_i$ is the ith change of the bond force and $R_j$ is the jth distance between the bond and AE hit centroid, and the centroid is determined as the geometrical center of the bond breakages. MTM ingeniously complies with this principle by monitoring all bond breakages during a simulation and treating bond breakages occurring close in time and space as a single hit. For the improvement of the operation efficiency, the single moment tensor of each AE hit is the maximum scalar-tensor, which can be determined by the eigenvalue of the moment tensor matrix ($m_j$):

$$M_0 = \left(\frac{\sum_{j=1}^{3} m_j^2}{2}\right)^{\frac{1}{2}} \tag{2}$$

The moment magnitude of the AE hit ($M_w$) can be determined with an empirical equation[42]:

$$M_w = \frac{2}{3}\log M_0 - 6 \tag{3}$$

Compared with the moment magnitude information, AE energy and *b*-value are more widely used to depict AE characteristics. It has been proved that the relationship between magnitude ($M_w$) and AE energy can be expressed by empirical formula Eq. (4)[43]:

$$\text{Log}E = 1.5 \times M_w + 4.8 \tag{4}$$

and the AE amplitude distribution conforms to the Gutenberg–Richter relationship:

$$\text{Log}N = a - b \times M_w \tag{5}$$

where *a* and *b* are constants, *N* is the number of simulated magnitudes that exceed $M_w$.

**2.2.2 AE analysis by MTM**

$PFC^{2D}$ was adopted to simulate the above experiment. The numerical model has a width of 50 mm and height of 100 mm which was made up of 16884 particles with the radius uniformly ranging from 0.21 mm to 0.35 mm. The parallel-bond model and smooth joint model were applied respectively to describe the mechanical response of the grain boundaries and joints of the actual shale specimen when loaded[44]. Upper and lower boundaries were applied to the relative velocity of ± 0.05m/s considering the amount of calculation[45]. Usually, the macro-scale mechanical behaviors of the rock relate to the micro properties, which should be calibrated by trial and error until simulated results are basically in line with the experimental ones[32]. In this study, the micro properties in Table 1 have been calibrated for a range of UCSs and Young's moduli based on the data presented in Table 2. Simulated AE parameters including AE hit, energy and *b*-value were acquired by MTM.

| Particle parameters | | Parallel bond parameters | |
|---|---|---|---|
| Density (kg/m$^3$) | 3000 | Bond effective modulus (GPa) | 22 |
| Effective modulus (GPa) | 22 | Bond stiffness ratio | 1.5 |
| Stiffness ratio | 1.5 | Tensile strength (MPa) | 170 |
| Friction coefficient | 0.8 | Cohesion (MPa) | 150 |
| Damping coefficient | 0.7 | Friction angle (°) | 40 |
| Smooth joint parameters | | | |
| Normal stiffness (GPa/m) | 10000 | Tensile strength (MPa) | 30 |
| Shear stiffness (GPa/m) | 3700 | Cohesion (MPa) | 80 |
| Friction coefficient | 20 | Joint friction angle (°) | 40 |

**Table 1.** Calibrated micro properties used in PFC$^{2D}$ to present the Marine shale specimen

| | Experiment | | | Simulation |
|---|---|---|---|---|
| | 1 | 2 | 3 | |
| UCS (MPa) | 188.16 | 198.42 | 166.67 | 173.12 |
| Young's modulus (GPa) | 32.34 | 27.37 | 23.78 | 27.59 |

**Table 2.** Calibrated results of macro properties

Figure 4 shows the simulated AE characteristics by MTM. During loaded with strain ranging from 0 to 0.36%, the stress-strain relation shows a linear manner while curves of cumulative energy and cracks basically keep invariant, which corresponds to few AE hits. These phenomena show that the synthetic rock model displays elastic property over this period. After that (strain from 0.36% to 0.65%), the stress curve fluctuates slightly while both curves of cumulative energy and cracks increase in the form of step-wise mode, involved in distinct AE hits. Finally, when the strain is over 0.65%, the stress-strain curve experiences in the postpeak stage, and curves of cumulative energy and cracks increase respectively from 0.0687 J to 2.5105 J and from 278 to 427 in number, corresponding *b*-value decreases from 1.06 to 0.83. These phenomena indicate the numerical model reaches the peak point.

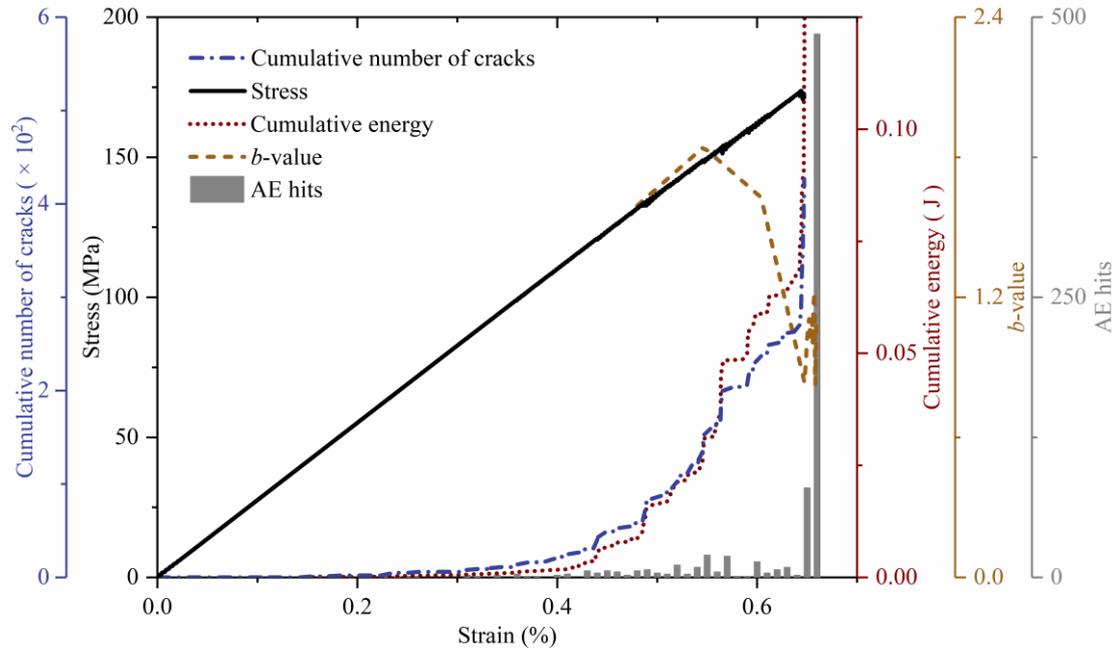

**Figure 4.** Axial stress, cumulative number of cracks and simulated AE parameters (cumulative energy, *b*-value and AE hits) versus axial strain by MTM

### 2.3 Comparison of AE characterization between MTM and the experiment

It is clear that AE characterization by MTM displays similar variation trends to the experimental results: at first there are few AE hits and invariability of energy and cracks, then AE hits increase and both curves of cumulative energy and cracks increase greatly, finally AE hits and curves of cumulative energy and cracks all occur breakthrough at the peak strength point, corresponding to the sharp decrease of *b*-values.

However, some discrepancies in AE characterization are remarkable. For the number of AE hits, the experiment records 30,502 while MTM records 734. Besides, at the beginning of loading, there are some AE hits by the experiment but not by MTM since AE hits caused by pre-existing defects such as natural cracks are difficult to be reproduced in simulation[46]. AE hits during the pre-peak stage are not obvious by MTM as well (Fig. 4). For the energy, comparisons show that energy by MTM is about 10 orders of magnitude more than that by the experiment. For the *b*-value, the simulated *b*-value acquired by MTM varies in a smaller range than that by the experiment. It is difficult to directly explain the above-mentioned discrepancies in AE characterization because MTM is inconsistent with experiments in principle, signal processing and energy analysis as illustrated in Fig. 1. To address these issues, it is necessary to fill in the gap between the experiment and MTM.

## 3   Particle-velocity-based model

### 3.1   Implementation

In order to fill in the gap between the experiment and MTM, a proxy model was proposed to imitate the experimental process and was used to compare with MTM. It used the particle velocity monitored to analyze AE characteristics of the model, named as particle-velocity-based model (PVBM).

As shown in Fig. 2c, the fracturing of the rock material is accompanied with the release of stored strain energy in the form of elastic waves. For the actual AE test, these elastic waves propagate inside the medium, then they can be detected by AE sensors attached to the sample[47]. The PVBM complies with the rules of the experiment at the most, that is, simulated AE signals are acquired by monitoring the normal particle velocity on a surface of a numerical model at each step in $PFC^{2D}$.

The next step is to analyze the simulated AE signals. In an actual AE test, AE signals are analyzed by a signal processor, of which a typical commercial software is the AEwin developed by Physical Acoustics Corporation (USA). However, the simulated AE signals cannot be analyzed directly by AEwin because the time evolution of $PFC^{2D}$ is computed via the step[39]. So a series of add-in code was developed by combining the principle of AEwin and the calculation characteristics of $PFC^{2D}$.

Finally, we acquired simulated AE characteristics of a rock model, mainly including hit, energy and *b*-value, which are designed in the following context (Fig. 5).

For the simulated hit, one or more predetermined evaluation thresholds should be set to identify simulated AE signals. When $|V(i)|$ exceeds the threshold, it marks the beginning of a simulated AE hit and $i$ is determined as an arrival step. As the step increases, the last threshold crossing is recorded as $V(i_L)$. Next, judge whether this simulated hit is over. The hit definition time (HDT) and Max duration are used to achieve this goal. If the threshold is not exceeded by $|V(i)|$ from $i_L$ to ($i_L$ + HDT), it marks the end. The other case is that a simulated AE signal will be cut off forcibly when its length reaches the Max duration, which is often used in the acquisition of continuous signals or the stage of very intense signals. After confirming the end of a simulated hit, other simulated AE features such as signal duration (*d*) and amplitude are recorded. To prevent reflections of the former signal from being taken as a start of the next signal, hit lockout time (HLT) is defined. Finally, $V(i + d + \text{HLT})$ will turn back to judge the next simulated hit until all data has been calculated.

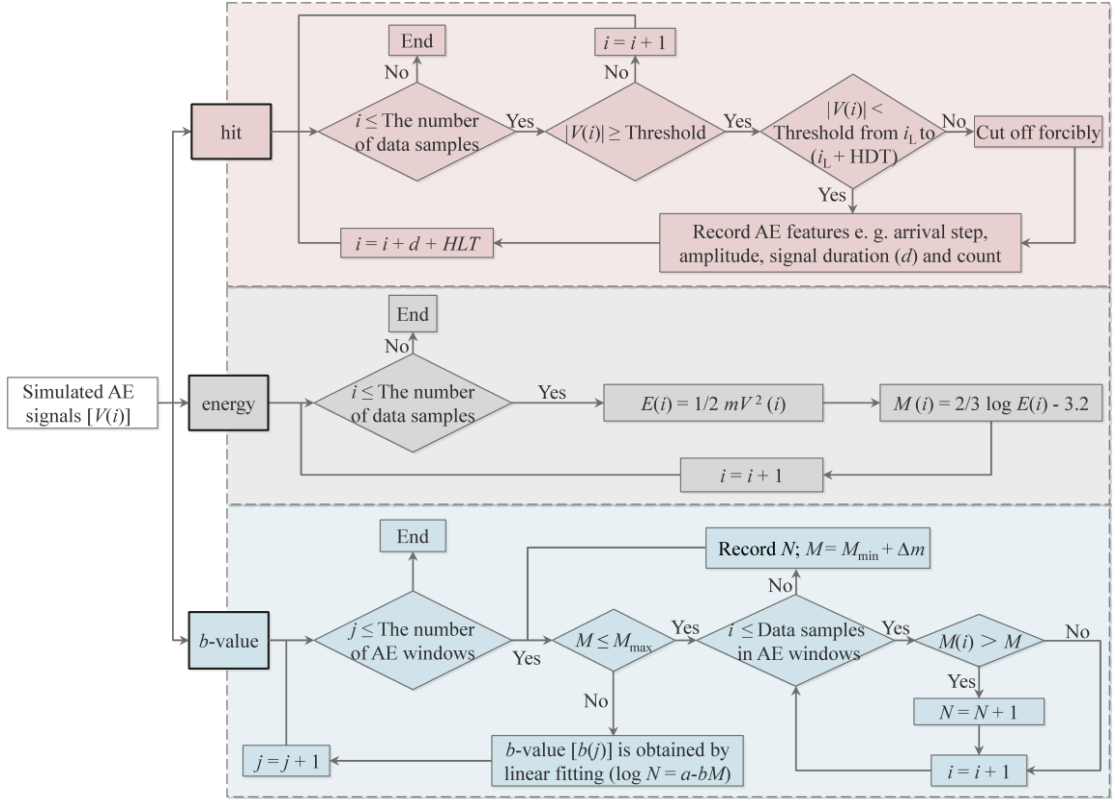

**Figure 5.** Flow chart of add-in code to calculate simulated AE characteristics. Simulated AE signal is represented by the normal velocity component $V(i)$ that is imported into three calculation modules to get the hit, energy and $b$-value, where $i$ is step rather than time

For the simulated energy, kinetic energy of particles can be determined explicitly in DEM. The simulated $b$-value can be calculated by Eq. (5). Reasonable magnitude range [$M_{min}$, $M_{max}$] and AE windows i.e. AE hit segmentation should be predetermined to achieve this linear fitting. For data samples $V(1) \sim V(i)$ in $j$th AE windows, $N$ is the number of simulated magnitudes that exceed ($M + \Delta m$), where $\Delta m$ is the magnitude interval. Then a set of correlation data between $N$ and $M$ is obtained to calculate $b(j)$ by Eq. (5) fitting with a least square method. After that, the $b(j + 1)$ will be calculated until all AE windows are completed.

### 3.2 AE characterization by PVBM
#### 3.2.1 Monitoring points
To keep the comparability, the numerical model and boundary conditions are the same as those in section

2.2.2 except that nine monitoring points a#~i# were set as shown in Fig. 6. Each monitoring point can be regarded as an AE sensor. They were equidistantly placed in the axial direction and lateral direction with a spacing of 20 mm and 25 mm respectively to receive normal velocity of monitoring points at each step. Then the acquired simulated AE signals were imported into the add-in code (Fig. 5) to get the final simulated AE parameters. Based on many trials and considering wane and wax, it is recommended that the evaluation threshold value, AE hit segmentation, Max duration, and HDT were set as 0.05 m/s, 50, 40 steps, and 20 steps, respectively.

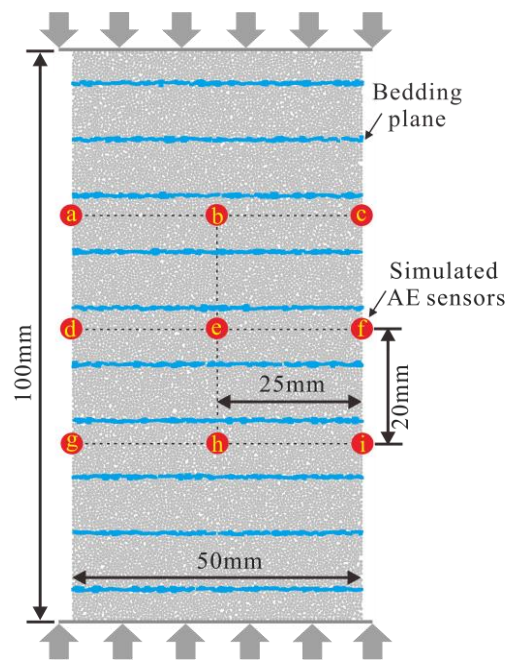

**Figure 6.** The layout of simulated AE sensors a#~i# on the numerical model. The axial and lateral spacing is 20mm and 25mm, respectively

### 3.2.2 Robustness of PVBM

To evaluate the robustness of PVBM, nine waveforms of simulated AE sensors a#~i# are all imported into the add-in code to get simulated AE parameters and to be compared.

Figure 7 shows the variation of simulated AE hits at AE sensors a#~i#. Results show a similar variation that the simulated AE hits emerge at the strain around 0.36% and experience several isolated peaks at the strain around 0.45%, 0.49%, 0.58% and 0.66% with increasing value in a sequence. These similarities show good robustness of AE hits by PVBM.

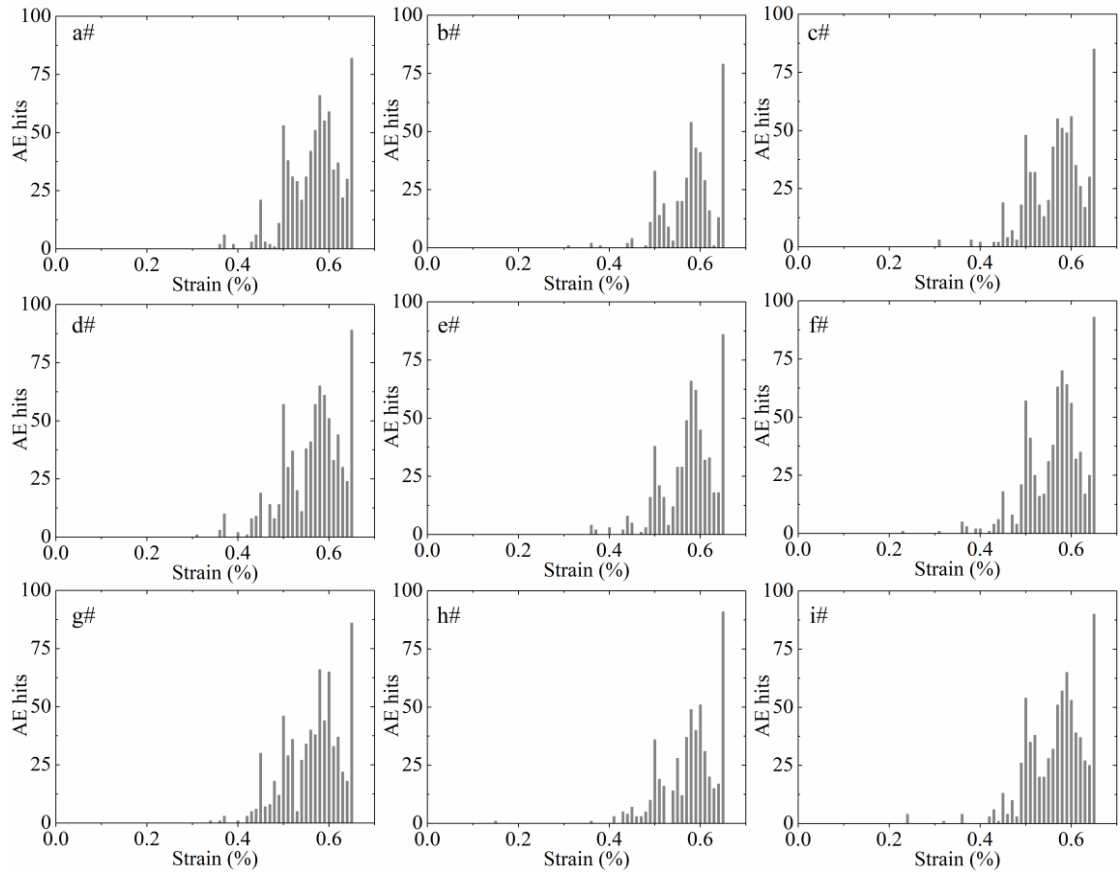

**Figure 7.** The number of simulated AE hits at AE sensors a#~i# versus axial strain by PVBM

Figure 8 shows the variation of simulated cumulative energy at AE sensors a#~i#. Results show a similar variation trend that the simulated cumulative energy increases firstly at the strain around 0.45%, then increases in the form of step-wise mode and finally soars at the strain around 0.64%. In this process, the strains corresponding to the thresholds of step-wise growth on the nine curves are almost the same at around 0.45%, 0.49%, 0.55% and 0.56%. In general, the identification of thresholds of cumulative AE energy by PVBM shows good robustness. Whereas, the variation magnitudes are a little different since some epicenters are away from the receiver but some are close.

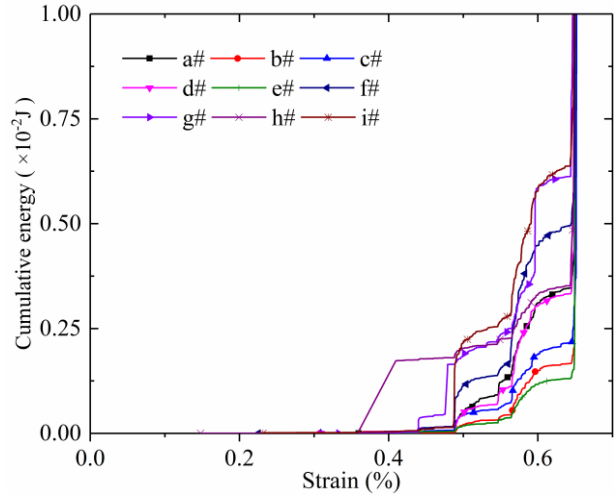

**Figure 8.** The variation of simulated cumulative energy at AE sensors a#~i# versus axial strain by PVBM

Figure 9 shows the variation of simulated *b*-value at AE sensors a#~i#. The trends of the simulated *b*-value display approximate hump curves. The *b*-values all begin to increase adjacent to the strain of 0.49% and 0.56%, indicating the increase of the proportion of fracturing events on a small scale. In contrast, the *b*-values all begin to decrease adjacent to the strain of 0.55% and 0.64%, and the latter fall more severely, corresponding to the peak point of the numerical model. Generally, the identification of turning points of *b*-values by PVBM displays good robustness.

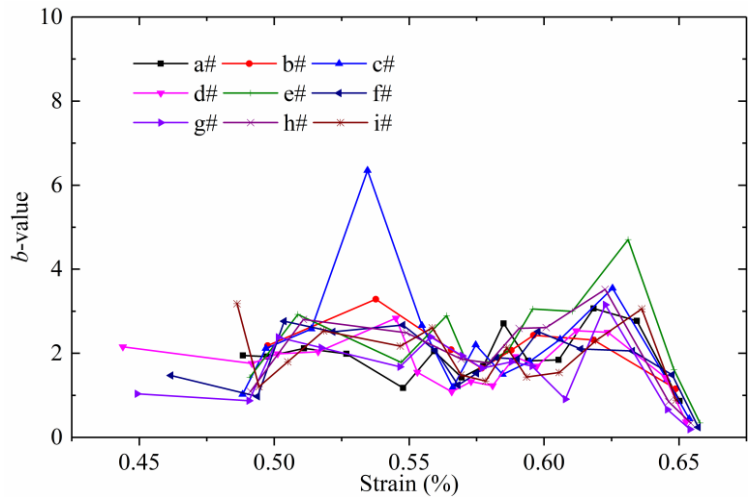

**Figure 9.** The variation of simulated *b*-value at AE sensors a#~i# versus axial strain by PVBM

AE characterization by PVBM shows good robustness. We used data at simulated AE sensor d# for systematical analysis. Considering the consistency of the numerical model and boundary conditions, the curves of stress and crack versus strain by PVBM are the same as those by MTM. As shown in Fig. 10, at first there are few AE hits and approximately invariant energy, then the former increases greatly and the latter increases in the form of step-wise mode, while the simulated *b*-value fluctuates in a hump manner. Interestingly, when the stress-strain curve fluctuates perceptibly, AE hits synchronously gather into solitary peaks, curves of cumulative energy and cracks concurrently increase in a step-wise manner, and the *b*-value simultaneously decreases, which indicates that there are fracturing events on a large scale at these thresholds, including at the strain around 0.45%, 0.55%, 0.59% and 0.65%[31]. The maximum extent is at the strain around 0.65% that AE hits increase from 24 to 89 and curves of cumulative energy and cracks soar respectively from 0.0034 J to 1.0625 J and from 278 to 427 in number, corresponding *b*-value decreases from 2.50 to 0.37. This indicates that the numerical model reaches the peak point.

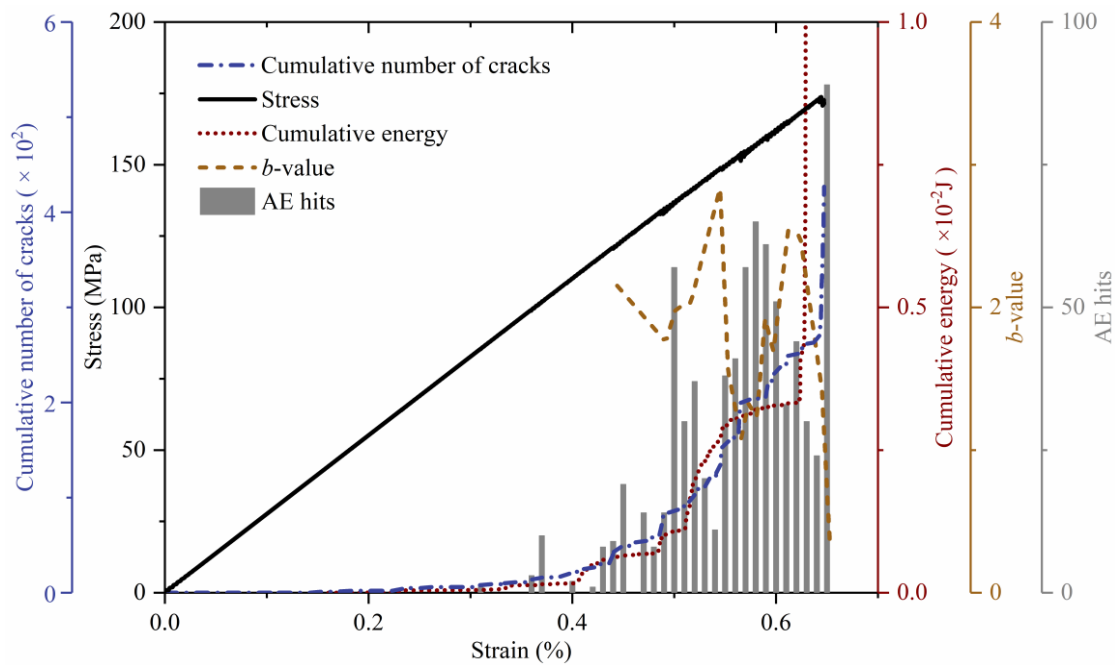

**Figure 10.** Axial stress, cumulative number of cracks and simulated AE parameters (cumulative energy, *b*-value and AE hits) versus axial strain at AE sensor d# by PVBM

## 4   Discussion

AE monitoring and modeling are very important for the prevention of rock mass engineering disasters. However, there are some contradictories between the actual AE test and AE simulation. By PVBM, we investigated these discrepancies and evaluated AE characterization including AE hit, energy and *b*-value of brittle rocks from the experiment to numerical simulation.

### 4.1   AE hit

Figure 11 shows the comparison of AE hits. The histograms all show a similar variation trend that AE hits are relatively few at first, then they increase and form approximate solitary peaks, before soaring at the peak points. However, this trend by MTM is not apparent compared with PVBM, manifested by smaller solitary peaks during the pre-peak stage and final excessive mutation at the peak point. It can be seen that the number AE hits of the pre-peak stage by MTM are about 2.5 times less than those by PVBM, while AE hits by MTM at the peak point are about 5.45 times more than those by PVBM.

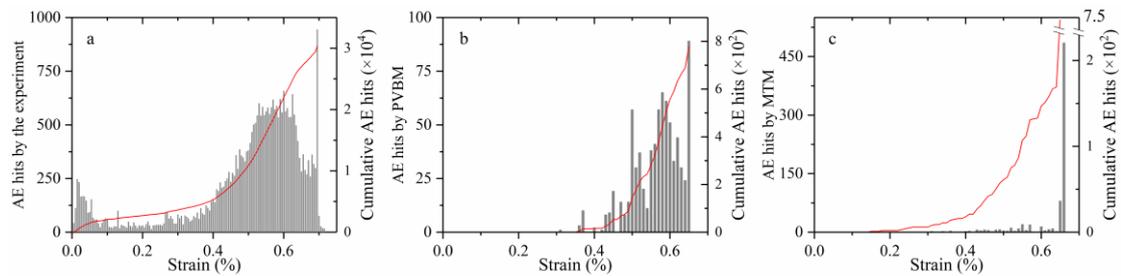

**Figure 11.** Comparison of AE hits by the (**a**) experiment, (**b**) PVBM and (**c**) MTM

For the difference of the pre-peak stage, one possible reason is that in MTM given that a newly-formed crack is within one particle diameter of an existing crack and the hit is still in the duration determined by assuming that a fracture propagates at half the shear wave velocity of the numerical model, this newly-formed crack and the existing crack will be considered as the same AE event[48]. The other possible reason is that there are signal reflections at interfaces when using PVBM, which causes the final received simulated AE signal to be a superposition of the wave from the epicenter and all its reflections[49].

For the difference at the peak point, since PVBM is based on waveform analysis, AE signals are difficult to be identified and separated accurately at the peak point as there are too many waveforms including the superposition and mutual influence of AE hits received in such a short period. These signals are always treated as continuous signals and are cut off forcibly[50]. Besides, the large-scale coalescence has a great influence on the transmission of elastic waves[51].

It is well known that the AE hits can intuitively reflect the progressive process of rock failure, thus it has been widely used for the prediction of rock mass engineering disasters. The above-mentioned evaluation of AE hits indicates that the pattern of AE hits by MTM is similar but not apparent compared with that by the experiment and PVBM, which hampers the identification of the pre-peak progressive fracturing.

Fortunately, PVBM covers this shortage. We thus recommend a combination of MTM and PVBM when using the numerical simulation to supplementarily predict rock mass engineering disasters.

**4.2 AE energy**

After the evaluations on AE hits, Fig. 12 shows the comparison of cumulative energy. The curves all show a similar variation trend that cumulative energy approximately keeps invariant at first, then increases greatly, and finally increases sharply and instantly at the peak point. Besides, during the pre-peak stages, both cumulative energy curves by PVBM and MTM perform in step-wise mode, corresponding thresholds are almost the same at the strain around 0.44%, 0.49%, 0.55%, 0.56%, 0.60% and 0.64%.

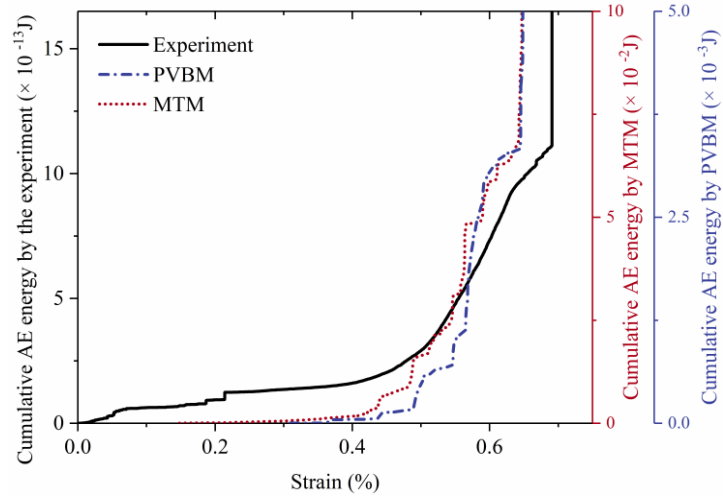

**Figure 12.** Comparison of cumulative energy by the experiment, PVBM and MTM.

However, the orders of magnitude of AE energy are different. AE energy by MTM is about $6\times10^{10}$ times more than that of the experiment, similar to the findings by Khazaei et al.[52]. Besides, AE energy by MTM is about 20 times more than that by PVBM. It is worthwhile to investigate the latter difference.

An intuitive explanation is the attenuation of elastic waves in the form of damping in DEM when using PVBM, which is necessary to reduce vibration by absorbing vibrational energy, manifested by the fact that low damping will lead to unnecessary continuous vibration, while high damping will lead to a decrease of AE amplitude and high-frequency content[39]. The present example is a quasi-static simulation with a local damping coefficient $\alpha = 0.7$, which has been proved to be high enough to prevent the formation of dynamic waves but insensitive to calculate AE energy by MTM[53]. Apart from the damping effect, radiated strain energy may be dissipated to friction at contacts and transfer to neighboring particles[19]. In this study, the AE energy by PVBM is only about 5% of that by MTM. It is very interesting since previous experimental studies coincidentally showed apparent energy (equivalently energy received at AE sensors in PVBM) was only 6% or less of the radiated energy (equivalently in-situ energy at the epicenter in MTM) of a hit[54].

There is a direct correlation between AE energy and the scale of the hazard. In the practical application, the actual AE test acquires energy after partially attenuation on the surface of the tested body, which is difficult to be calibrated since the invisibility of the fracturing events. While PVBM allows this calibration since the fracturing events are visible and analyzable in numerical simulation. Besides, MTM can explicitly calculate the in-situ energy, which is helpful for the evaluation of the scale of the potential hazard. In order to reduce potential losses at the most, the practical AE test is suggested to be combined with both PVBM and MTM.

### 4.3  *b*-value

Figure 13 shows the comparison of the *b*-value. The three curves have a similar variation trend that at first fluctuating and then decreasing significantly at the peak points. However, the valid strain range to calculate the *b*-value and its magnitude fluctuation are different. As for the difference in valid strain range, the *b*-value is respectively calculated from the strain around 0.03%, 0.44%, and 0.48% by the experiment, PVBM and MTM, since there is no AE hit at the beginning of the simulation. As for the difference in magnitude fluctuation, the lower magnitude fluctuation of the *b*-value by MTM is mainly resulted from the narrower range of magnitudes according to a previous study[55]. Actually, some researchers questioned the reliability of the *b*-values because it was calculated in a statistical method by Eq. (5) to describe AE amplitude distribution and was susceptible to the researcher's subjectivity such as the selections of the number of AE hits, AE hit segmentation and AE amplitude range[56]. For MTM, the variation of the number of AE hits is consistent with that of cracks because cracks close in time and space are treated as a single AE hit. It is natural to connect AE hits with the number of bonds, which is directly related to the particle size. Khazaei et al.[34] also pointed out that the number of AE hits was a function of model resolution and found that coarser particles would result in smaller *b*-values.

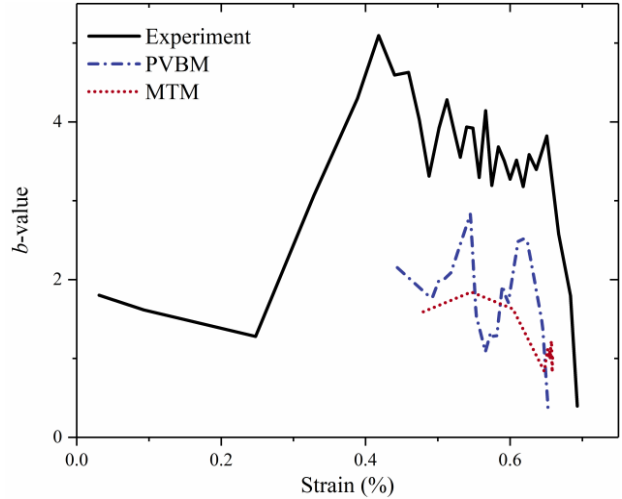

**Figure 13.** Comparison of *b*-value by the experiment, PVBM and MTM

It has been proved that the *b*-value can identify states of damage despite the limitation in accurately evaluating the degree of damage[33]. As shown in Fig. 13, *b*-value curves of both PVBM and MTM increase when the strain falls into the range (0.49% - 0.55%), corresponding to the formation of new cracks and restriction of crack propagation. Then they decrease at the strain around 0.55%, which indicates fracturing events on a relatively large scale. However, there remains the difference of which the subsequent *b*-value by PVBM increases but *b*-value by MTM decreases continuously. This would be explained as multiple reflections of elastic waves induced by previous fracturing events on a large scale. Finally, *b*-values by MTM and PVBM both decrease at the related peak point.

The *b*-value is always adopted to characterize the scale of AE magnitude distribution. The identification of damage states by the *b*-value plays an important role in the prevention of rock mass engineering disasters. The above comparison indicates that the errors of experimental *b*-values resulted from the undesirable noise in the actual AE test can be refrained from numerical simulation. Considering both the reliable prevention and low cost in engineering practice, the actual AE test should be combined with PVBM and MTM to accurately identify damage states.

### 4.4 Consistency and compatibility

As a proxy model, PVBM is highly consistent with the experiment in principle and process. Besides, PVBM is compatible with MTM since both are implemented in the same numerical model. Comparisons indicate that AE characterization by PVBM shows consistent variation trends with those by the experiment and MTM, especially AE hits by PVBM is closer to that by the experiment and the step-wise mode of cumulative energy curve by PVBM is closer to that by MTM. Furthermore, compared with MTM, PVBM has advantages of smaller computation amount and easier programming, reflected by about 5 times higher calculation speed in the present simulation. However, PVBM is not as accurate as MTM since it inherits some limitations from experiments such as reflection and attenuation of elastic waves. Hence, PVBM may be used as an alternative model in qualitative AE characterization for some very complex problems such as particle-intensive models and special geometrical structures.

Although MTM is not compatible with the experiment in principle and process, the feasibility was verified via PVBM. Comparisons indicate that AE characterization by MTM shows high consistency with PVBM in AE energy. In addition, MTM is of great accuracy since it is not involved in the influence of reflection and attenuation of elastic waves, which indicates that MTM is applicable to the ideal case. However, there is no perfect model[57]. MTM has high time consumption, especially for a 3D fine model with small particles.

### 4.5    New insights into future AE characterization and applications

According to the current study, the AE experiment is realistic but constrained by attenuation and reflection of elastic waves. Compared with the experiment, MTM can calculate in-situ energy more explicitly, while it is difficult to apparently identify the pre-peak progressive fracturing. Nevertheless, this limitation of MTM can be made up by PVBM, which is a proxy model to imitate the experimental process and is not as accurate as MTM. Besides, both PVBM and MTM can reduce errors in the identification of damage states. Therefore, we can acquire new insights into future AE characterization and applications: only by combining the AE experiment with both PVBM and MTM can engineers acquire AE characterization more reasonably and accurately, which provides improvements in the protection and prevention of rock mass engineering disasters.

## 5    Conclusion

Comparison of AE characterization between MTM and the experiment revealed that there were some remarkable discrepancies between them, including principle, processing method and energy analysis. In order to fill in these gaps, this paper proposed a proxy named PVBM to imitate the experimental process. PVBM provided a reasonable evaluation on AE characterization from the experiment to numerical simulation. It was found that the experiment, PVBM and MTM were all able to quantitatively describe

AE characterization in the progressive failure process of the brittle rock, acquired AE parameters including AE hit, energy and *b*-value all showed similar variation patterns. While the AE experiment acquired actual AE characterization but had limitations such as attenuation and reflection of elastic waves. MTM accurately calculated the AE energy but could not apparently characterize the variation trend of AE hits during the pre-peak stages, which respectively indicated the improvement in the evaluation on hazard scale but the instability in the identification of hazard precursor. With good robustness, PVBM was consistent with the experiment in principle and process and compatible with MTM, which was used to reasonably explain the quantitative differences of AE characterization between the experiment and MTM. AE hits by PVBM were closer to that by the experiment, while AE energy by PVBM showed excellent consistency at thresholds corresponding to the step-wise growth stage with that by MTM. Besides, both PVBM and MVM reduced experimental *b*-value errors in the identification of damage states. In the practical AE application, it was suggested that a systematic combination of the advantages of PVBM and MTM could effectively prevent rock mass engineering disasters.

## Acknowledgements

This work was financially supported by the National Key Research and Development Program of China under Grant No. 2019YFC1509701, the National Natural Science Foundation of China under Grant Nos. 41977249, 42090052, and U1704243, and the Science Foundation of Key Laboratory of Shale Gas and Geoengineering, Institute of Geology and Geophysics, Chinese Academy of Sciences under Grant No. KLSG201709.